\documentclass[aps,prl,floatfix,epsfig,twocolumn,showpacs,preprintnumbers]{revtex4}
\usepackage{amsmath}
\usepackage{graphicx}
\usepackage[version=3]{mhchem}
\usepackage{dcolumn}
\usepackage{multirow}
\usepackage{bm}
\usepackage{subfigure}
\usepackage{float}
\usepackage[T1]{fontenc}
\usepackage[colorlinks=true,linkcolor=blue,urlcolor=blue,citecolor=blue]{hyperref}

\begin{document}

\title{Moir\'{e} effects in graphene--hBN heterostructures}
\author{Yongping Du$^{1,2,3}$, Ning Xu$^{4}$, Xianqing Lin$^{5}$, and
Antti-Pekka Jauho$^{1,2}$}
\thanks{Corresponding author: antti@dtu.dk}
\affiliation{$^{1}$Center for Nanostructured Graphene, Technical University of Denmark,
Kongens Lyngby, Denmark\\
$^{2}$DTU Physics, Technical University of Denmark, Kongens Lyngby, Denmark\\
$^{3}$Department of Applied Physics and Institution of Energy and
Microstructure, Nanjing University of Science and Technology, Nanjing
210094, China\\
$^{4}$Department of Physics, Yancheng Institute of Technology, Yancheng
224051, China\\
$^{5}$College of Science, Zhejiang University of Technology, Hangzhou
310023, China}

\begin{abstract}
Encapsulating graphene in hexagonal Boron Nitride has several advantages:
the highest mobilities reported to date are achieved in this way, and
precise nanostructuring of graphene becomes feasible through the protective
hBN layers. Nevertheless, subtle effects may arise due to the differing
lattice constants of graphene and hBN, and due to the twist angle between
the graphene and hBN lattices. Here, we use a recently developed model which
allows us to perform band structure and magnetotransport calculations of
such structures, and show that with a proper account of the moir\'{e}
physics an excellent agreement with experiments can be achieved, even for
complicated structures such as disordered graphene, or antidot lattices on a
monolayer hBN with a relative twist angle. Calculations of this kind are
essential to a quantitative modeling of twistronic devices.
\end{abstract}

\maketitle
\date{\today }

\section{Introduction}

Graphene, the first successfully isolated two-dimensional material, has
opened a new hot research area\cite{Graphene-1,Graphene-2}. Due to the
linear bands crossing the Fermi level, low-energy carriers in graphene
behave like massless, relativistic Dirac fermions, allowing predictions from
quantum electrodynamics to be tested in a solid-state system. The high Fermi
velocity\cite{highVF}, Dirac-cone band structure\cite{Graphene-1} and
ultra-strong mechanical properties\cite{mechanical} make graphene a
promising material for next-generation electronic nanodevices and high-speed
switching devices. However, the intrinsic zero energy gap of graphene has
hampered its applications in modern electronics. In a practical
nanoelectronic device semiconducting graphene is necessary.

A sizable band gap opening around the Fermi level in a graphene antidot
lattice (GAL, a regular arrangement of antidots in a graphene lattice) has
been predicted by several theoretical studies\cite%
{GAL-PRL,GAL-theory-2,GAL-theory-3,GAL-theory-4,GAL-theory-5}, and was
recently realized in an experiment\cite{Nat-tech-2019}. The band gap in GAL
can be tuned by the size, shape, and symmetry of both the antidot and the
superlattice cell\cite%
{GAL-PRL,GAL-theory-2,GAL-theory-3,GAL-theory-4,GAL-theory-5}.
The tunable band gap can be used to design quantum wells and channels for
electronic devices\cite%
{GAL-PRL,GAL-theory-2,GAL-theory-3,GAL-theory-4,GAL-theory-5}.
Interestingly, transport under magnetic fields in an antidot lattice is
predicted to show Hofstadter butterfly features arising from the competition
between the antidot lattice periodicity and the magnetic length\cite%
{GAL-magnetic}.

Recently, heterostructures consisting of graphene and hexagonal boron
nitride (G/hBN) have drawn intense attention\cite%
{Nat-tech-2019,G-hBN-PRL-relax,G-hBN-PRL-relax-2,G-hBN-Hofstadter-PRL,G-hBN-Hofstadter,G-hBN-Hofstadter-sci,G-hBN-Cloning-Dirac,G-hBN-NC-band-gap,G-hBN-PRL-gap,G-hBN-FQH-sci,G-hBN-review,G-hBN-review-2,G-hBN-natphys}%
. The lattice mismatch between graphene and hBN causes moir\'{e} patterns
with long wavelengths to emerge when graphene and hBN lattices are exactly
aligned or twisted relatively by a small angle\cite%
{G-hBN-PRL-relax,G-hBN-PRL-relax-2}. 
Experiments have revealed many exciting phenomena, such as the Hofstadter
butterfly\cite%
{Nat-tech-2019,G-hBN-Hofstadter,G-hBN-Hofstadter-sci,G-hBN-Cloning-Dirac,G-hBN-review,G-hBN-review-2}
(now arising from the competition between the lattice mismatch induced moir{%
\'{e}} length and the magnetic length), or the fractional quantum Hall effect%
\cite{G-hBN-FQH-sci,G-hBN-review,G-hBN-review-2}. However, the twist-angle
dependence of the properties of antidot lattices defined on G/hBN
heterostructures has not yet been the subject of a systematical experimental
or theoretical study. Another system where moir{\'{e}} effects show up
dramatically is twisted bilayer graphene, where unconventional
superconductivity or correlated insulator behavior may occur at certain
twist angles between the monolayers \cite{BiGr-Nature1,BiGr-Nature2}.

In this paper we consider two examples of recent experimental interest where
the relative angle between graphene and hBN plays an important role\cite%
{Nat-tech-2019,Expdisorder} 
: (i) disordered graphene, and (ii) antidot lattices. We first summarize our
most important physical findings, and discuss the technical details in
subsequent paragraphs. By using an effective lattice model, we first
calculate the electronic structure and conductance of G/hBN with and without
a relaxation of the graphene and hBN monolayers comprising the system.
Our results show that without relaxation the band structure is particle-hole
symmetric, in disagreement with experimental data, while the fully relaxed
graphene shows, correctly, a particle-hole asymmetry emphasizing the
importance of lattice relaxation of G/hBN. Next, we compute the conductance
in the presence of a magnetic field perpendicular to the graphene sheet. The
computed magnetoconductance shows a moir\'{e} potential induced secondary
Dirac point which clones the Landau fan of the primary Dirac point. The
Hofstadter butterfly features are also observed in our numerical results.
The computed results are in excellent agreement with experimental data\cite%
{Nat-tech-2019,G-hBN-Hofstadter,G-hBN-Hofstadter-sci,G-hBN-Cloning-Dirac}.
Based on an Anderson model, we next investigate how disorder affects the
electronic structure and conductance of G/hBN. We find that even though
disorder can lift the degeneracy of the bands at high symmetry points, the
main features of the band structure are similar to the clean case. While the
magnitude of the conductance with disorder is reduced to almost a half of
the one without disorder, the magnetoconductance stays unchanged. Finally,
we systematically study the electronic structure and transport behavior of
antidot lattice on twisted G/hBN. A major theoretical finding is that the
secondary Dirac point will disappear once the distance between antidot edges
("the neck width", denoted by $d_{n}$ and see Appendix F) is smaller than the moir\'{e} wave length, a feature seen
in experiments \cite{Nat-tech-2019}.

\section{The model}

Our effective tight-binding model is proposed in Ref.\cite{model-PRB} and we
follow the procedures outlined there (Appendix A summarizes the details pertinent to this work). Carbon atoms can be removed
from the antidot regions by removing the associated rows and columns from
the system Hamiltonian. Any dangling $\sigma $ bonds for a carbon atom with
only two neighboring carbon atoms are assumed to be passivated with Hydrogen
atoms so that the $\pi $ bands are unaffected. Transport quantities are
calculated using recursive Green's function techniques, following Ref.\cite%
{greenfunction}. The zero-temperature conductance is given by the Landauer
formula $G=\frac{2e^{2}}{h}T$, where $T$ is transmission coefficient. A
finite magnetic field $B$ perpendicular to the graphene plane is modeled by
associating a Peierls phase to the hopping amplitude $t\rightarrow
te^{i\varphi _{ij}}$, where $\varphi _{ij}=(e/\hbar )\int_{r_{i}}^{r_{j}}%
\mathbf{A}\cdot d\mathbf{r}$. Here $\mathbf{A}$ is the vector potential and $%
r_{i}$ is the position of atom $i$. We choose the Landau gauge $\mathbf{A}%
=(-By,0,0)$, and the Peierls phase becomes $\varphi _{ij}=\frac{eB}{\hbar }%
(x_{j}-x_{i})(\frac{y_{i}+y_{j}}{2})$. In the leads the magnetic field is
set to zero.


Microscopic theoretical investigations on the G/hBN system are made
cumbersome by the size of the unit cell, which for small relative rotations
contains thousands of atoms making first-principles calculations very
expensive. Effective continuum models\cite{continmodel} or several
tight-binding models\cite{G-hBN-Hofstadter-PRL,empirical} with empirical
parameters controlling the interlayer interaction between graphene and hBN
have been applied to calculate the electronic properties of G/hBN. However,
the effective continuum model cannot be used to simulate the transport for
realistic experimental conditions, while results for the tight-binding
models with empirical parameters must be carefully scrutinized to ascertain
their reliability. For a large device transport simulation Chen et al.\cite%
{scalerpotential} applied a scaled graphene lattice with a triangular
periodic scalar moir\'{e} potential, and successfully reproduced the main
features of the secondary Dirac point. However, as we show below, this
simple moir\'{e} potential does not lead to particle-hole asymmetry. In our
effective lattice model, the Hamiltonian terms at any local part of a
twisted and relaxed G/hBN only depend on the local relative shift and
relaxation-induced strain and can be derived from a transparent set of
parameters calculated by density functional theory. Moreover, our effective
lattice model can be used to calculate the electronic structure of G/hBN
with \textit{any} twist angle, and does not require a recalculation of the
parameters for a new twist angle.

\begin{figure}[tbh]
\center\includegraphics[scale=0.5]{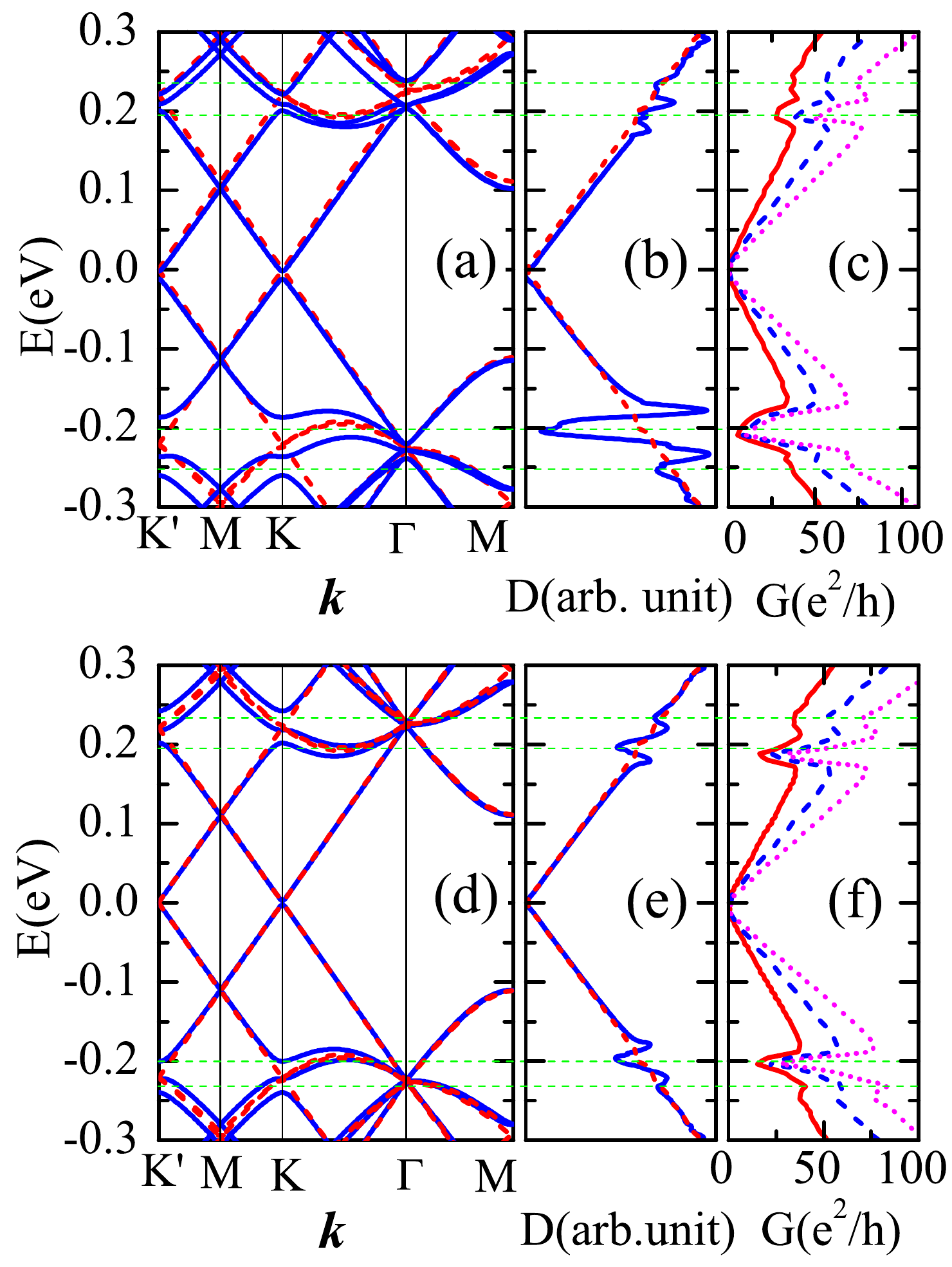}
\caption{The calculated electronic structure of G/hBN with twist angle 1.0047$^{\circ }$.
(a), (b) and (c) are calculated with relaxation while (d), (e),
and (f) are without lattice relaxation. (a) and (d) show the band
structures; (b) and (e) are the densities of states; (c) and (f) give the
conductance without magnetic field. In the band structure and DOS, the red
dashed lines denote the graphene monolayer for comparison. In the
conductances (c) and (f), the red solid, blue dashed, and magenta dot lines
are calculated for device sizes 200 nm$\times$ 200 nm, 300 nm$\times$300 nm,
and 400 nm$\times$400 nm, respectively. The thin dashed green lines indicate
the minima in DOS.}
\label{Fig1}
\end{figure}

We next calculate the band structure and conductance for the twist angle $%
\theta $=1.0047$^{\circ }$ with and without lattice relaxation; the results
are reported in Fig. \ref{Fig1}. The band structure (Fig.\ref{Fig1}(a)) and density of states
(DOS) (Fig.\ref{Fig1}(b)) of fully relaxed G/hBN show a particle-hole
asymmetry which is consistent with the experiments\cite%
{Nat-tech-2019,G-hBN-Hofstadter,G-hBN-Hofstadter-sci,G-hBN-Cloning-Dirac,G-hBN-PRL-gap,G-hBN-FQH-sci,G-hBN-review,G-hBN-review-2}%
. To emphasize the importance of full relaxation, we plot in Fig. \ref{Fig1}%
(d) the band structure of graphene with a scalar moir{\'{e}} potential\cite%
{scalerpotential}: the band degeneracy at high symmetry points is lifted and
one finds secondary Dirac points at $\pm $0.225eV. However, in this case the
DOS around the two secondary Dirac points are equal and obey particle-hole
symmetry (Fig.\ref{Fig1}(e)), in contrast to experiment, and Fig. \ref{Fig1}%
(a) and \ref{Fig1}(b). The reason for the difference is that a fully relaxed
graphene lattice has, in addition to a modified on-site energy, also a
modified hopping between neighboring C atoms.


Appendices B and C report the band structure and
conductance for additional twist angles and edge orientations. The two main
conclusions are: (i) the secondary Dirac points shift to larger energies as
the twist angle increases, because the interaction between graphene and hBN
decreases as the twist angle increases (see Fig. \ref{Figs1}-\ref{BS-141}), and (ii) the
calculated conductance does not significantly depend on whether the edges of
the simulated device are in the armchair or zigzag directions (see Fig. \ref{Figs3}).
In subsequent calculations we consider a device with zigzag edges.

\begin{figure}[tbh]
\center\includegraphics[scale=0.8]{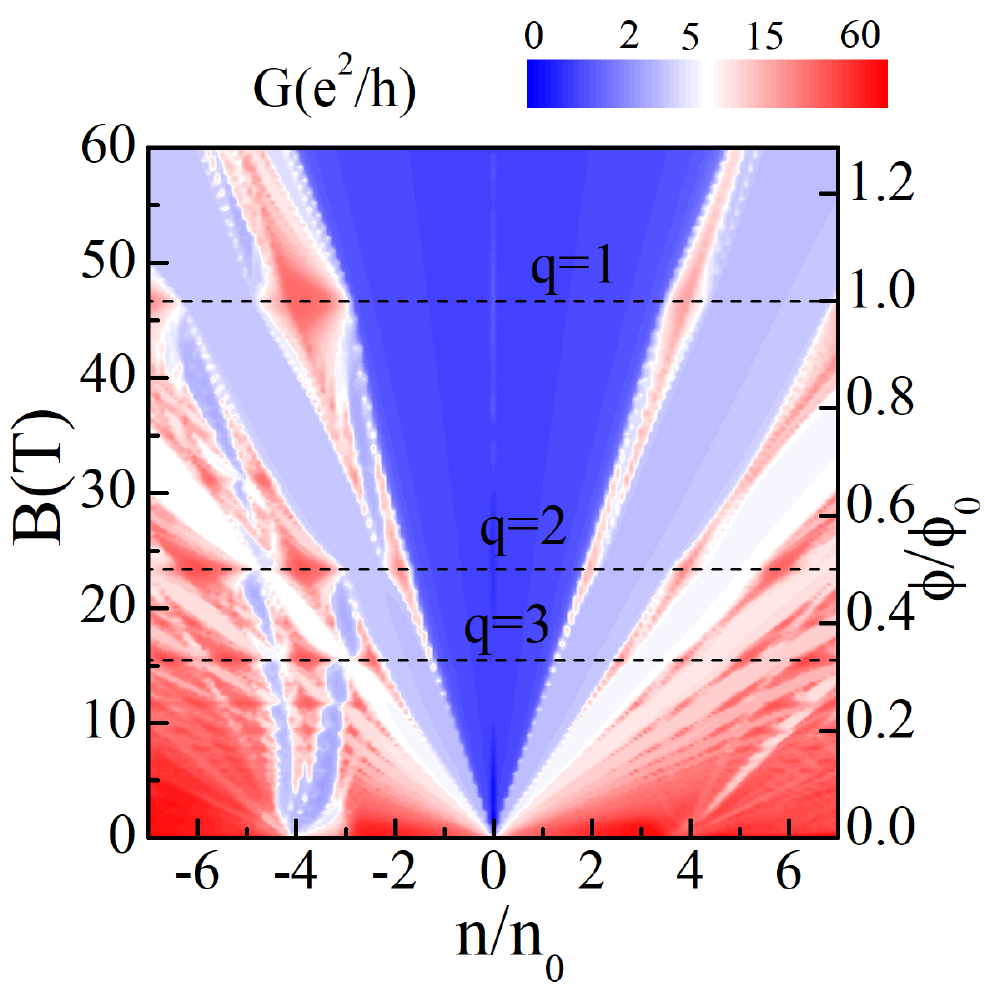}
\caption{Longitudinal magnetoconductance $G(B,n)$ as a function of magnetic
field and electron density. Here $n$ is electron density, $n_{0}=1/S$ is the electron density per each
Bloch band, where $S$ is the area of supercell. Right vertical axis is scaled by $\protect\phi /\protect\phi %
_{0} $ where $\protect\phi $ is the flux through one moir{\'{e}} unit cell
and $\protect\phi _{0}=h/e$ is the flux quantum. Dashed black lines show $q$%
=1, 2, and 3, where the Landau levels intersect.}
\label{fig2}
\end{figure}


\begin{figure*}[tbh]
\centering
\includegraphics[scale=0.6]{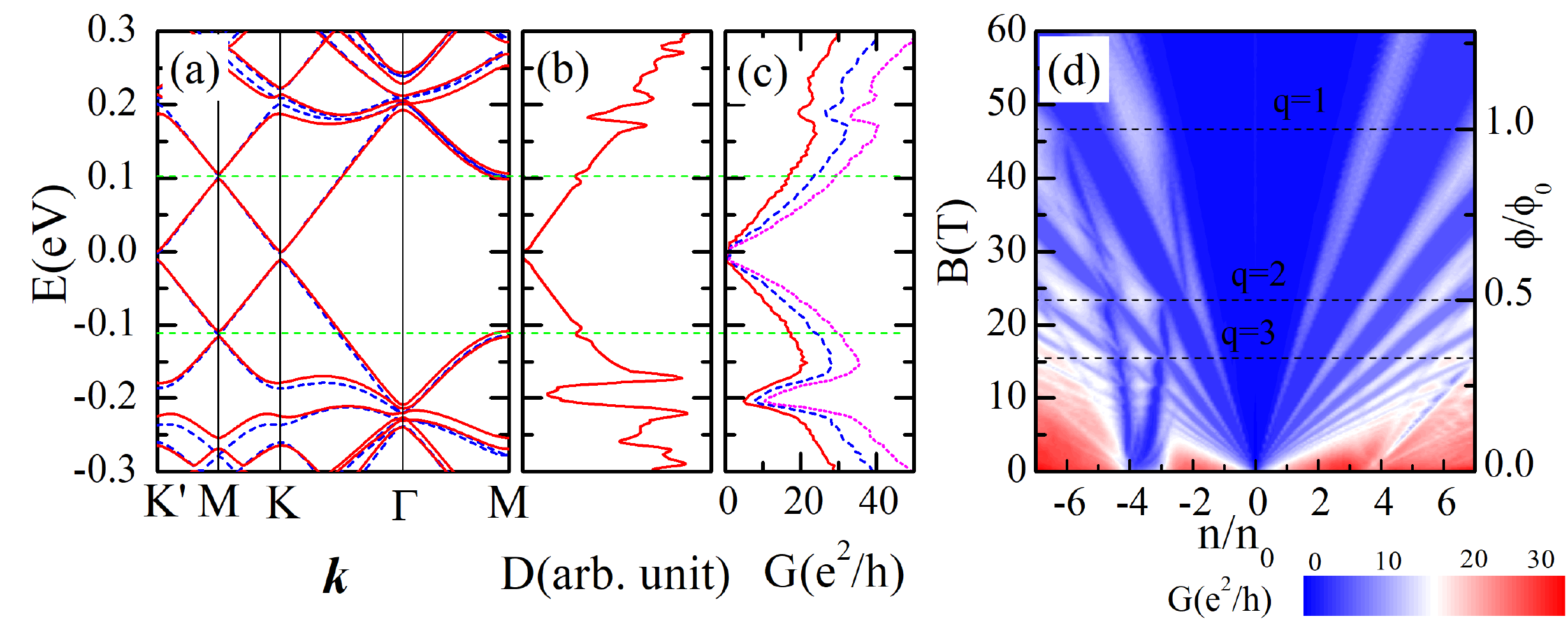}
\caption{The electronic structure and conductance of G/hBN with disorder
with twist angle 1.0047$^{\circ }$. (a) Red solid line: band structure with
disorder. Dashed blue line: band structure for pristine sample. (b) Density
of states. (c) Zero-field conductance. Red, blue, and magenta lines refer to
device sizes of 200 nm $\times$ 200 nm, 300nm $\times$ 300 nm and 400nm $%
\times$ 400nm, respectively. (d) Magnetoconductance for a disordered sample.}
\label{Fig3}
\end{figure*}

We next carry out magnetotransport simulations using the effective lattice
model for a 300 nm$\times $300 nm device ($\simeq 10^{7}$ atoms).
Our results are shown in Fig. \ref{fig2} (some finer details in $G$, not
visible in Fig. \ref{fig2}, are discussed in Appendix D). Both the primary ($%
n/n_{0}=0$) and the secondary Dirac points ($n/n_{0}=\pm 4$) break into
sequences of Landau levels upon application of a magnetic field, forming the
so called Hofstadter butterfly spectrum. The high conductance areas (red,
white) separate the gapped Landau levels (blue).
As mentioned above, the moir\'{e} superlattice potential breaks the
partical-hole symmetry, as also seen in transport experiments \cite%
{Nat-tech-2019,G-hBN-Hofstadter,G-hBN-Hofstadter-sci,G-hBN-Cloning-Dirac,G-hBN-PRL-gap,G-hBN-FQH-sci,G-hBN-review,G-hBN-review-2}%
. 
As the magnetic field increases, the Landau levels will intersect when $%
\frac{\phi }{\phi _{0}}=\frac{1}{q}$, where
$q$ is an integer, indicated in Fig. \ref{fig2} with black dashed lines. The
intersection of the primary and secondary Landau levels leads to a closing
of the magnetic band gap, and a resulting high conductivity, seen as bright
spots in Fig. \ref{fig2}.
Results for other twist angles, showing the same general trends, are given
in Appendix B.

\section{Effects due to disorder}
Disorder is ubiquitous to all graphene samples, even for those synthesized
with state-of-the-art technologies\cite{disorder-1}. The properties of
nanoribbons are known to be strongly affected by disorder \cite%
{disorderribbon}, and recent studies suggest that the electronic and
transport properties of graphene antidot lattice may also be strongly
perturbed by relatively modest disorder\cite%
{disorder-2,disorder-3,disorder-4,disorder-5,disorder-6}. An exploration of
the effect of disorder on G/hBN is thus called for. Here, we introduce
disorder as a site-diagonal random potential with matrix element $%
H_{ij}=\delta _{ij}v_{i}$, where $v_{i}$ are independent, uniformly
distributed random variables in the range of [-$V_{0}$,$V_{0}$] (where $%
V_{0} $ is set to 0.5eV larger than the onsite energy $\varepsilon _{i}$,
maximum 0.14eV), and have zero mean and unit variance. Other details on the
disorder model are provided in Appendix E. 
One would expect that disorder breaks certain symmetries with concomitant
modifications in the band structure. Here, our simulations show that
disorder leads to band degeneracy lifting at high symmetry points,
especially it opens a band gap at the M point (Fig. \ref{Fig3}(a)). The band
gap opening leads to a kink in DOS (indicated by horizontal green lines in
Fig. \ref{Fig3} (b)). Even though the DOS is modified by disorder, the
generic features in conductance stay qualitatively unchanged, except for an
overall reduction of $\simeq 50\%$ in magnitude (compare Fig. \ref{Fig1}(c)
and \ref{Fig3}(c)). 
In particular, the features in the conductivity at the secondary Dirac
points still remain. Overall, we conclude that the transport properties of
G/hBN at $B=0$ are very robust against disorder.

To investigate whether the robustness persists for finite magnetic fields,
we next calculate the magnetotransport properties of disordered G/hBN; the
results are shown in Fig. \ref{Fig3}(d). One observes that the main features
of the magnetoconductivity are essentially the same as those without
disorder, shown in Fig. \ref{fig2}. The main differences are the overall
reduction of magnetoconductiviy, as already discussed above, and that many
fine features are washed out by disorder, and thus the results for the
disordered system appear significantly more regular than those for the
pristine system. The conclusion is that the magnetotransport in G/hBN is
indeed robust with respect to disorder, and its salient features should
survive even a nonideal fabrication process.


\begin{figure}[H]
\includegraphics[scale=0.7]{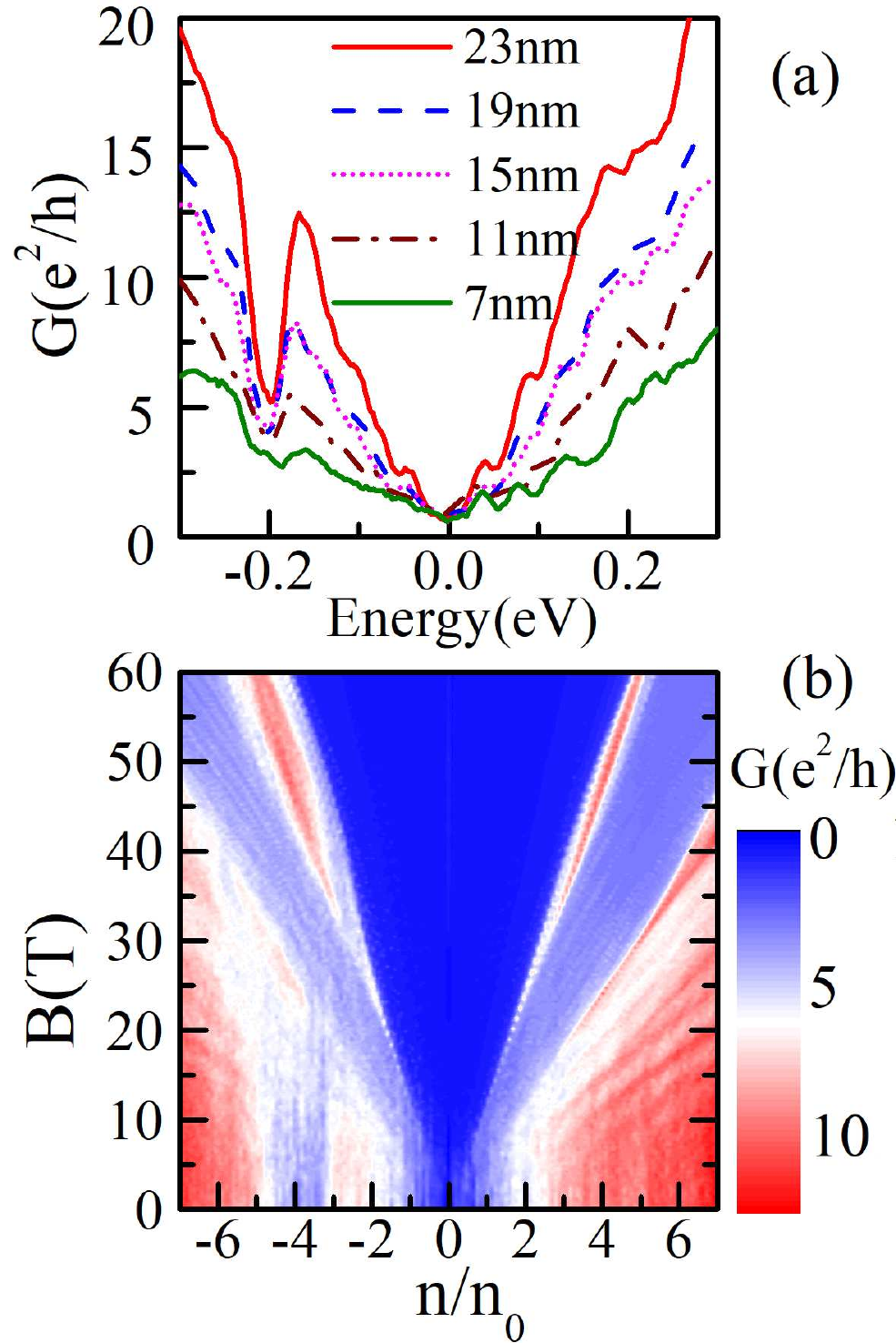} \centering
\caption{(a) Zero-field conductances for antidot lattices with different
neck widths ($d_{n}=7$, $11$, $15$, $19$, $23$ nm). The secondary feature at
$E\simeq -0.2$ eV disappears if the neck width $d_{n}$ is smaller than moir%
\'{e} length $\protect\lambda =10.1$ nm for the twist angle=1.0047$^{\circ}$%
. (b) Landau fan diagram with for neck width $d_{n}=15$ nm.}
\label{Fig4}
\end{figure}

\section{Antidot lattices}
The band structure of graphene antidot lattices (GAL) may differ
qualitatively from that of graphene, as witnessed by the observation of a
band gap in recent experiments\cite{Nat-tech-2019}. 
We next describe how the techniques developed in this work can yield
additional important information of transport in a GAL on a (twisted) hBN
substrate. In addition to the magnetic length, there are now two (at least)
other competing other length scales: the moir{\'{e}} length $\lambda $,
which is maximally 14.4 nm\cite{model-PRB}, and the length scale(s)
characterizing the GAL.
A schematic geometry of the GAL system is shown in Fig. \ref{Figs7}. 
We consider triangular antidot lattices, because they are known to lead to
gapped systems\cite{GAL-theory-4} 
Our numerical results reveal an important relationship between $\lambda $
and $d_{n}$. The secondary Dirac point features in conductance are observed
only when $d_{n}>\lambda $, while they vanish if $d_{n}<\lambda $ as shown
in Fig. \ref{Fig4}(a). This finding explains an important detail in the
recent experiments\cite{Nat-tech-2019}. Namely, pristine graphene
encapsulated in hBN shows a primary Landau fan and \textit{two} cloned fans
corresponding to moir{\'{e}} periods of 10.2nm and 17.2nm. The second moir%
\'{e} length is larger than the longest possible moir\'{e} length in single
layer graphene on hBN. Recent experimental and theoretical studies show that
the second moir\'{e} pattern is due to the simultaneous effects of top and
bottom hBN \cite{supermoire-1,supermoire-2,supermoire-3,supermoire-4}.
However, after fabrication of the GAL the second peak related to the 17.2 nm
moir\'{e} wavelength is lost. As the neck length in the fabricated GALs is
12-15nm, and thereby smaller than the second moir\'{e} wavelength 17.2 nm,
one does not expect to see Landau fans related to this length scale, which
indeed is the case in the experiment. In Fig. \ref{Fig4}(b) we show the
Landau fan diagram of G/hBN with an antidot lattice whose $d_{n}=15$ nm.
Compared to the results shown in Fig. \ref{fig2}, there is a reduction in
both the magnitude of magnetoconductance, and the number of Landau levels.
Importantly, the secondary Dirac point survives, just as in experiments\cite%
{Nat-tech-2019}.

\section{Conclusion}

In summary, we have performed a systematic examination of the consequences
of the lattice mismatch and relative orientation between graphene and hBN,
and show that several experimental observations find a common explanation
rooted in the interplay of the moir{\'e} length, and other relevant length
scales. The method described in this work can be extended to many other
systems of current interest, including twistronic devices.

\section{Acknowledgments}

The work was supported by the Jiangsu Province Science Foundation for Youth
(Grant No. BK20170821), National Science Foundation of China for Youth
(Grant No. 11804160) and National Natural Science Foundation of China (Grant
No. 11974312). The Center for Nanostructured Graphene (CNG) is sponsored by
the Danish National Research Foundation, Project No. DNRF103.

\appendix
\renewcommand\thetable{A\arabic{table}}
\renewcommand\thefigure{A\arabic{figure}}
\section{Appendix A: Effective lattice model}
\setcounter{table}{0}
\setcounter{figure}{0}
The effective lattice Hamiltonian for graphene and hexagonal boron nitride
(G/hBN) reads
\begin{widetext}
\begin{equation}
H=\sum_{i}(h_{Ai}c_{Ai}^{\dagger }c_{Ai}+h_{Bi}c_{Bi}^{\dagger
}c_{Bi})-\sum_{<i,j>}t_{ij}(c_{Ai}^{\dagger }c_{Bj}+c_{Bj}^{\dagger }c_{Ai}),
\label{Hef}
\end{equation}%
\end{widetext}
where $c_{mi}^{\dagger }$ is the creation and $c_{mi}$ is the annihilation
operator of $p_{z}$ state in sublattice $m$ and unit cell $i$, $h_{mi}$ and $%
t_{ij}$ represent on-site energies and hopping terms. Here we set $%
h_{0}=(h_{A}+h_{B})/2$ and $h_{z}=(h_{A}-h_{B})/2$. The hopping terms along
the three nearest-neighbor vectors $\delta _{n}$ with $n=1-3$ (see Fig.1 in
Ref.\cite{model-PRB}) are represented by $-t_{n}$.

\subsection{First step: hopping terms and on-site energies at point $%
\mathbf{r}$ for rigid G/hBN}

As the analysis in Ref.\cite{model-PRB} shows, in a large moir\'{e}
superlattice the local lattice structure is similar to that of a shifted
G/hBN bilayer where both layers have the same orientation and the same
lattice constant as monolayer graphene (see Fig. 1 in Ref.\cite{model-PRB}). First-principles calculations show that the electronic structure of
the shifted G/hBN is periodic in the shift vector $\mathbf{d}$ with the
lattice vectors of graphene determining the period. Thus, for every shift
vector $\mathbf{d}$, we can derive the hopping parameters $t_{n}$ and
on-site energies $h_{0}$ and $h_{z}$ by fitting the band structure from
first-principles calculations. The resluts show that $t_{n}$, $h_{0}$ and $%
h_{z}$ are also periodic in the shift vector $\mathbf{d}$. Since the $t_{n}$%
, $h_{0}$ and $h_{z}$ in effective lattice model are periodic in vector $%
\mathbf{d}$, they can be expanded in Fourier series, such as
\begin{eqnarray}
t_{n}(\mathbf{d}) &=&\sum_{\mathbf{G}}\widetilde{t}_{n}(\mathbf{G})\cos [%
\mathbf{G}\cdot \mathbf{d}+\phi _{n}(\mathbf{G})]  \label{tnd} \\
h_{0}(\mathbf{d}) &=&\sum_{\mathbf{G}}\widetilde{h}_{0}(\mathbf{G})\cos [%
\mathbf{G}\cdot \mathbf{d}+\phi _{0}(\mathbf{G})]  \label{h0} \\
h_{z}(\mathbf{d}) &=&\sum_{\mathbf{G}}\widetilde{h}_{z}(\mathbf{G})\cos [%
\mathbf{G}\cdot \mathbf{d}+\phi _{z}(\mathbf{G})]  \label{hz}
\end{eqnarray}%
where $\mathbf{G}$ are the reciprocal lattice vectors of graphene and 10
shortest vectors including the origin are used in the expansion. $\overset{%
\sim }{t}_{n}(\mathbf{G})$ ($\overset{\sim }{h}_{0}(\mathbf{G})$, $\overset{%
\sim }{h}_{z}(\mathbf{G})$) and $\phi _{n}(\mathbf{G})$ ($\phi _{0}(\mathbf{G%
})$, $\phi _{z}(\mathbf{G})$)are the amplitudes and corresponding phase. The
expansion parameters for $t_{n}$, $h_{0}$ and $h_{z}$ are listed in Table %
\ref{Table1}.

\begin{table*}[tbh]
\centering%
\begin{tabular}{ccccccccccc}
\hline\hline
$G$ & $\widetilde{t}_{1}$ & $\phi _{1}(^{\circ })$ & $\widetilde{t}_{2}$ & $%
\phi _{2}(^{\circ })$ & $\widetilde{t}_{3}$ & $\phi _{3}(^{\circ })$ & $%
\widetilde{h}_{0}$ & $\phi _{0}(^{\circ })$ & $\widetilde{h}_{z}$ & $\phi
_{z}(^{\circ })$ \\ \hline
(0,0) & 2540.23 & 0.00 & 2540.23 & 0.00 & 2540.23 & 0.00 & 0.00 & 0.00 & 2.13
& 0.00 \\
(1,0) & 16.78 & 132.90 & 10.45 & -151.18 & 16.78 & 132.90 & 25.12 & -80.83 &
12.47 & 179.52 \\
(-1,1) & 16.78 & 132.90 & 16.78 & 132.90 & 10.45 & -151.18 & 25.12 & -80.83
& 12.47 & 179.52 \\
(0,-1) & 10.45 & -151.18 & 16.78 & 132.90 & 16.78 & 132.90 & 25.12 & -80.83
& 12.47 & 179.52 \\
(1,1) & 2.55 & -17.69 & 2.55 & 17.69 & 4.05 & 0.00 & 2.86 & -180.00 & 1.29 &
0.00 \\
(-2,1) & 4.05 & 0.00 & 2.55 & -17.69 & 2.55 & 17.69 & 2.86 & -180.00 & 1.29
& 0.00 \\
(1,-2) & 2.55 & 17.69 & 4.05 & 0.00 & 2.55 & -17.69 & 2.86 & -180.00 & 1.29
& 0.00 \\
(2,0) & 2.20 & -103.20 & 0.98 & 60.31 & 2.20 & -103.20 & 1.95 & 51.47 & 1.04
& -52.14 \\
(-2,2) & 2.20 & -103.20 & 2.20 & -103.20 & 0.98 & -60.31 & 1.95 & 51.47 &
1.04 & -52.14 \\
(0,-2) & 0.98 & -60.31 & 2.20 & -103.20 & 2.20 & -103.20 & 1.95 & 51.47 &
1.04 & -52.14 \\ \hline\hline
\end{tabular}%
\caption{The parameter of the Fourier expansion of $t_{n}(n=1-3)$, $h_{0}$,
and $h_{z}$. $\protect\widetilde{t}_{n}$, $\protect\widetilde{h}_{0}$, and $%
\protect\widetilde{h}_{z}$ are the expansion amplitudes and are in the unit
of meV. $\protect\phi _{n}$, $\protect\phi _{0}$, and $\protect\phi _{z}$
are the corresponding expansion angles. $\mathbf{G}=n_{1}\mathbf{b}_{1}+n_{2}%
\mathbf{b}_{2}$ are the used reciprocal lattice. $\mathbf{b}_{1}$ and $%
\mathbf{b}_{2}$ are the reciprocal vector for graphene. The data is adapted
with permission from Ref.\protect\cite{model-PRB}.}
\label{Table1}
\end{table*}

With the obtained parameters, we can get the hopping terms and on-site
energies around a point \textbf{r} for effective lattice model because the
local lattice structure of moir\'{e} superlattice can be approximated as a
shifted bilayer with $\mathbf{d}(\mathbf{r})$. For a rigid superlattices, $%
\mathbf{d}(\mathbf{r})=(S^{-1}-I)\mathbf{r}$, where $I$ is unit matrix and $%
S=\frac{%
\begin{array}{c}
1%
\end{array}%
}{%
\begin{array}{c}
1+\epsilon%
\end{array}%
}\left(
\begin{array}{cc}
\cos \theta & -\sin \theta \\
\sin \theta & \cos \theta%
\end{array}%
\right) $, $\theta $ is twist angle between graphene and hexagonal boron
nitride (hBN), $\epsilon =(a_{G}-a_{hBN})/a_{hBN}$. Here, $a_{G}$ and $%
a_{hBN}$ are the lattice constant for graphene and hBN, respectively.

\subsection{Second step: relaxation of G/hBN}

Due to the energy gain from the larger domains of energetically favorable
stacking configurations (AB stacking), the rigid G/hBN bilayer undergoes
spontaneous relaxation. The full lattice relaxation can be calculated by
solving three equations self-consistently\cite{model-PRB} as we show
below. We define $\mathbf{u}^{1}(\mathbf{r})$ and $\mathbf{u}^{2}(\mathbf{r})
$ (see Fig. 1 in Ref\cite{model-PRB}, and Fig. 3 in \cite{BilayerG})%
\textbf{\ }are the displacement vector for top graphene layer and bottom hBN layer,
respectively. The total energy ($E_{tot}$) of bilayer supercell is the
summation of the elastic energy ($E_{el}$) and interlayer interaction energy
$E_{int}$ and also is a functional of the displacement vector $\mathbf{u}%
^{j}(\mathbf{r})$. The elastic energy ($E_{el}$) is given by\cite{model-PRB,BilayerG}:

\begin{widetext}
\begin{equation}
E_{el}=\overset{2}{\sum_{\mathbf{j=1}}}\int d\mathbf{r}\left\{ \frac{\lambda
_{j}+\mu _{j}}{2}\left( \mathbf{\frac{\partial u_{x}^{j}}{\partial x}+\frac{%
\partial u_{y}^{j}}{\partial y}}\right) ^{2}+\frac{\mu _{j}}{2}\left[ \left(
\mathbf{\frac{\partial u_{x}^{j}}{\partial x}-\frac{\partial u_{y}^{j}}{%
\partial y}}\right) ^{2}+\left( \mathbf{\frac{\partial u_{y}^{j}}{\partial x}%
+\frac{\partial u_{x}^{j}}{\partial y}}\right) ^{2}\right] \right\}
\label{Eel}
\end{equation}%
\end{widetext}

where $\lambda _{1}=3.653eV/\mathring{A}^{2}$, $\mu _{1}=9.125eV/\mathring{A}%
^{2}$; $\lambda _{2}=1.779eV/\mathring{A}^{2}$, $\mu _{2}=7.939eV/\mathring{A%
}^{2}$ are the elastic Lam\'{e} factors for graphene and hBN, respectively.

The interlayer interaction energy $E_{int}$ can be written as\cite{model-PRB,BilayerG}:

\begin{equation}
E_{int}=\int V[\mathbf{d}(\mathbf{r})]d\mathbf{r}  \label{Eint1}
\end{equation}%
where $V[\mathbf{d}(\mathbf{r})]=\widetilde{V}\underset{k=1}{\overset{3}{%
\sum }}\cos (\mathbf{G}_{k} \cdot \mathbf{d}(\mathbf{r})+\phi _{V})$. For $%
\mathbf{G}_{k}$ we have $\mathbf{G}_{1}=\mathbf{b}_{1}$, $\mathbf{G}_{2}=-%
\mathbf{b}_{1}+\mathbf{b}_{2}$, $\mathbf{G}_{3}=-\mathbf{b}_{2}$. $%
\widetilde{V}$ takes 4.38 while $\phi _{V}=-50.26%
{{}^\circ}%
$.

The minimization of total energy ($E_{tot}$) as a functional of $\mathbf{u}%
^{j}(\mathbf{r})$ leads to a set of Euler-Lagrange equations by using a
similar procedure as in Ref.\cite{model-PRB,BilayerG}:

\begin{widetext}
\begin{equation}
\left(
\begin{array}{c}
\widetilde{u}_{x}^{j}(\mathbf{q}) \\
\widetilde{u}_{y}^{j}(\mathbf{q})%
\end{array}%
\right) =(-1)^{j}\left(
\begin{array}{cc}
(\lambda _{j}+2\mu _{j})q_{x}^{2}+\mu _{j}q_{y}^{2} & (\lambda _{j}+\mu
_{j})q_{x}q_{y} \\
(\lambda _{j}+\mu _{j})q_{x}q_{y} & (\lambda _{j}+2\mu _{j})q_{y}^{2}+\mu
_{j}q_{x}^{2}%
\end{array}%
\right) ^{-1}\left(
\begin{array}{c}
f_{x}^{j}(\mathbf{q}) \\
f_{y}^{j}(\mathbf{q})%
\end{array}%
\right)  \label{EL1}
\end{equation}
\end{widetext}

where $\mathbf{q}$ takes each $\mathbf{G}_{s}$, $\mathbf{G}_{s}$ are the
reciprocal lattice vectors of the supercell and 60 shortest nonzero ones
have been used. $j=1,2$ represent top graphene layer ($j=1$) and bottom hBN
layer ($j=2$). The Fourier components $\widetilde{\mathbf{u}}^{j}(\mathbf{G}%
_{s})$, and $\mathbf{f}^{j}(\mathbf{G}_{s})$ are defined as

\begin{widetext}
\begin{eqnarray}
\mathbf{u}^{j}(\mathbf{r}) &=&\sum_{\mathbf{G}_{s}}\widetilde{\mathbf{u}}%
^{j}(\mathbf{G}_{s})e^{i\mathbf{G}_{s}\cdot \mathbf{r}}  \label{self1} \\
\frac{\partial V(\mathbf{d})}{\partial \mathbf{d}} &=&-\widetilde{V}\underset%
{k=1}{\overset{3}{\sum }}\sin (\mathbf{G}_{k}\cdot \mathbf{((S^{-1}-I)%
\mathbf{r+u}^{\mathbf{1}}\mathbf{(r)-u}^{2}\mathbf{(r)})}+\phi _{V})\mathbf{G%
}_{k}=\sum_{\mathbf{G}_{s}}\mathbf{f}^{j}(\mathbf{G}_{s})e^{i\mathbf{G}%
_{s}\cdot \mathbf{r}}  \label{s2}
\end{eqnarray}
\end{widetext}
Equations (\ref{EL1}), (\ref{self1}), and (\ref{s2}) can be solved
self-consistently to obtain converged $\widetilde{\mathbf{u}}^{j}(\mathbf{G}%
_{s})$, when the difference of $\triangle \widetilde{\mathbf{u}}^{j}(\mathbf{%
G}_{s})=\left( \left\vert \widetilde{\mathbf{u}}^{j}(\mathbf{G}_{s})_{n}-%
\widetilde{\mathbf{u}}^{j}(\mathbf{G}_{s})_{n-1}\right\vert \right) ^{2}$
between two steps is less than 10$^{-5}$. After relaxation, one can then get
the displacement vector $\mathbf{u}^{j}(\mathbf{r})$ and at last the
modified shift vector $\mathbf{d(r)=(S^{-1}-I)\mathbf{r+u}^{\mathbf{1}}%
\mathbf{(r)-u}^{2}\mathbf{(r)}}$ being used to calculate hopping terms and
on-site energies.

\subsection{Third step: the strain effect on hopping terms and
on-site energies}

Fully relaxed graphene is under strain, which will influence the hopping
terms and on-site energies. Here we use the formulation of the dependence of
Hamiltonian parameters on the strain tensor proposed by Fang \textit{et al}%
\cite{strain}. At a position $\mathbf{r}$, the changes in hopping terms ($%
\delta t_{n}$) and on-site energies ($\delta h_{i}$) can be expressed as

\begin{widetext}
\begin{eqnarray}
\delta t_{n} &=&\alpha (\frac{\partial u_{x}^{1}}{\partial x}+\frac{\partial
u_{y}^{1}}{\partial y})+\beta \lbrack \widehat{\delta }_{ny}(\frac{\partial
u_{x}^{1}}{\partial x}-\frac{\partial u_{y}^{1}}{\partial y})+\widehat{%
\delta }_{nx}(\frac{\partial u_{x}^{1}}{\partial y}+\frac{\partial u_{y}^{1}%
}{\partial x})],  \label{strain1} \\
\delta h_{i} &=&\alpha _{0}(\frac{\partial u_{x}^{1}}{\partial x}+\frac{%
\partial u_{y}^{1}}{\partial y})  \label{strain2}
\end{eqnarray}
\end{widetext}

where $n=1-3$, $i=A$ and $B$, the strain is in the graphene layer, and $%
\widehat{\delta }_{n}$ is the unit vector along $\delta _{n}$. Based on
first-priciples calculations of strained graphene, the parameter $\alpha $, $%
\beta $, and $\alpha _{0}$ are fitted to be 3.27eV, -4.40eV and -4.95eV,
respectively.
\begin{widetext}

\appendix
\renewcommand\thetable{B\arabic{table}}
\renewcommand\thefigure{B\arabic{figure}}
\section{Appendix B: Additional twist angles}
\setcounter{table}{0}
\setcounter{figure}{0}

\begin{figure}[H]
\centering
\includegraphics[scale=0.5]{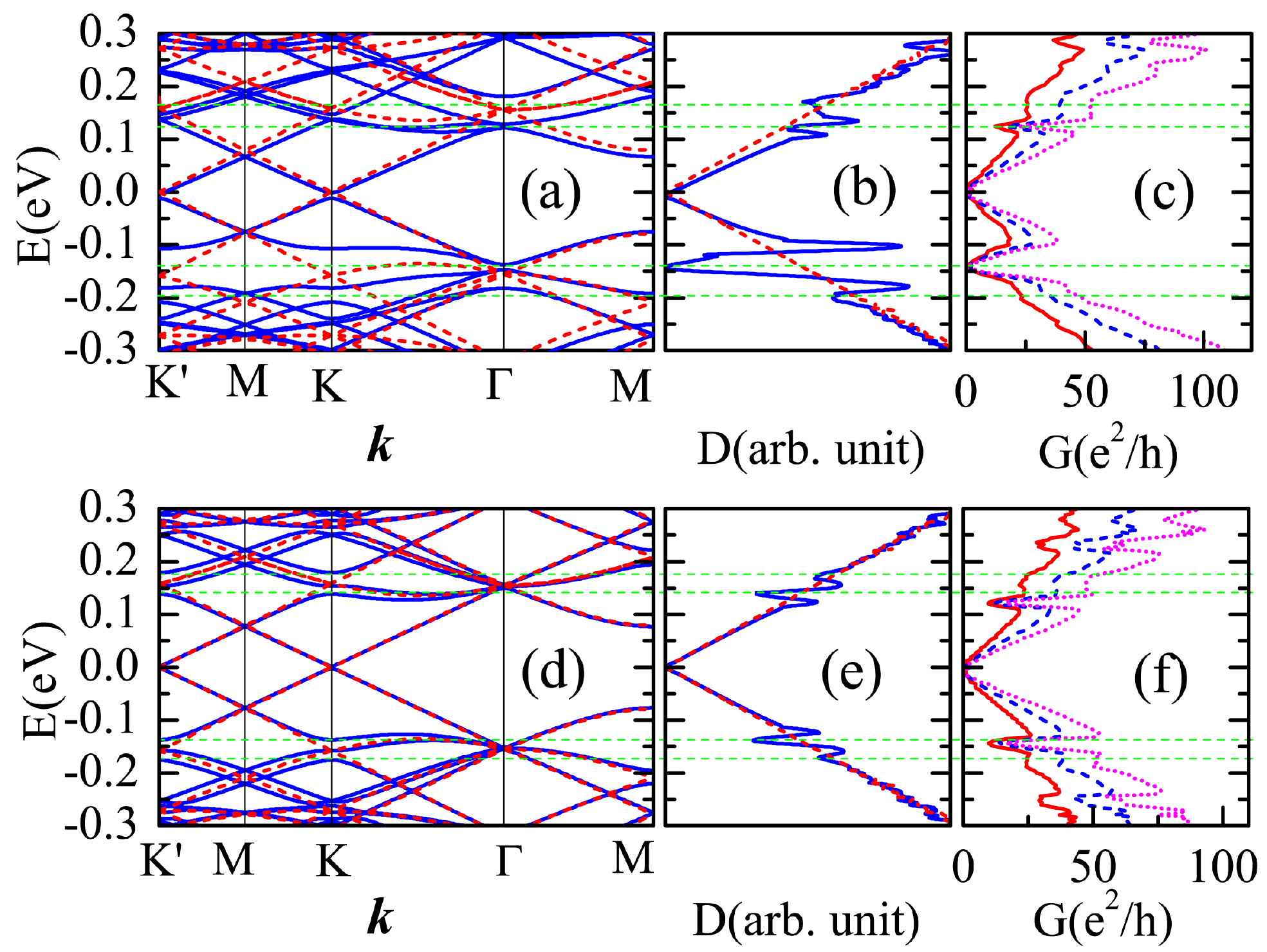}
\center\caption{The calculated electronic structure of G/hBN with twist angle 0.0$%
^{\circ}$. (a), (b) and (c) are calculated with relaxation while (d), (e),
and (f) are without lattice relaxation. (a) and (d) display the band
structures; (b) and (e) are the densities of states; (c) and (f) are the
conductance without magnetic field. In the band structure and DOS, the red
dashed lines denote the graphene monolayer results. In the conductance
(c) and (f), the red solid, blue dashed, and magenta dot lines are calculated
for device sizes 200nm$\times $200nm, 300nm$\times $300nm, and 400nm$\times $%
400nm, repectively.}
\label{Figs1}
\end{figure}

\begin{figure}[H]
\centering
\center\includegraphics[scale=0.5]{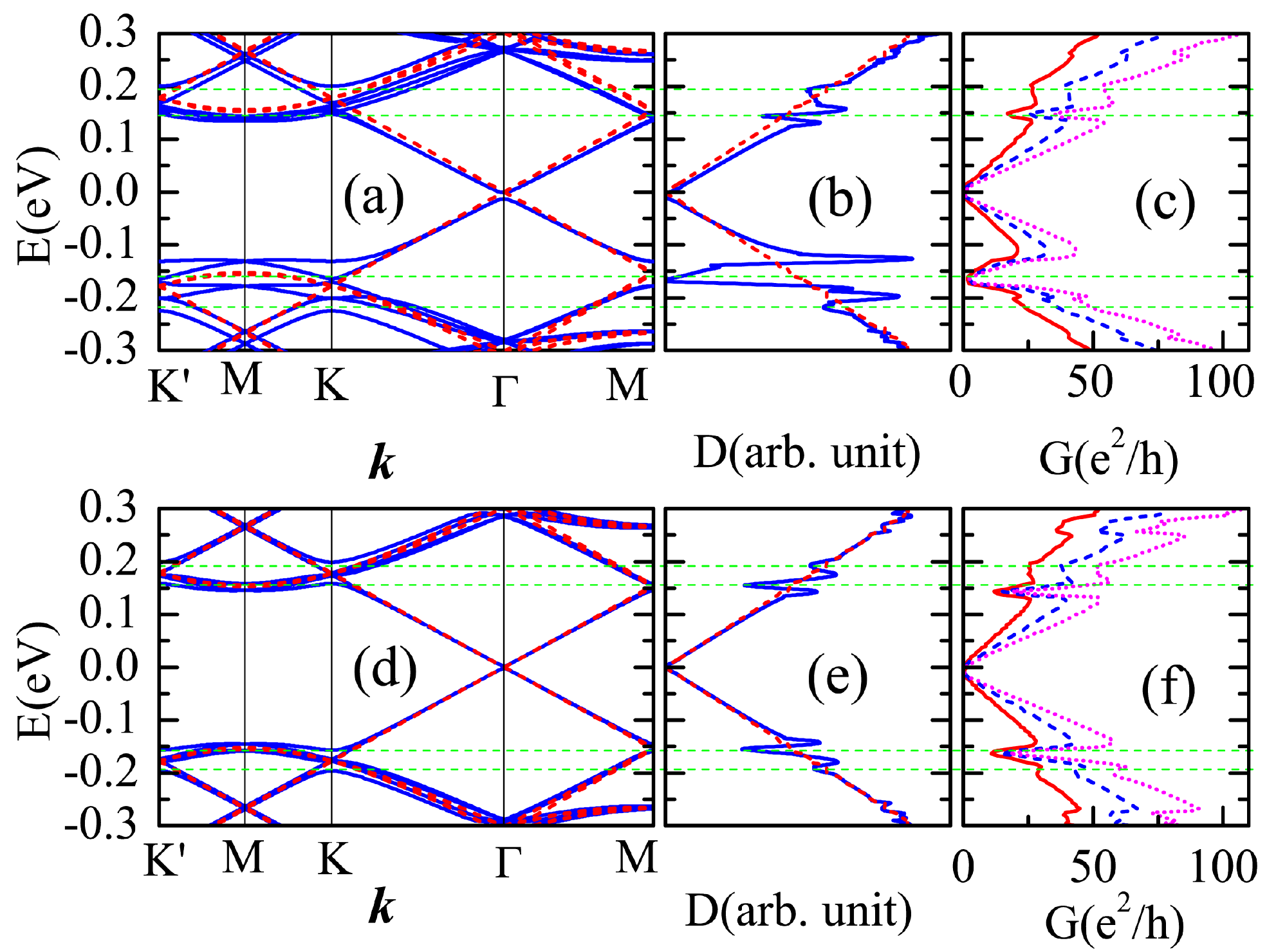}
\center\caption{The calculated electronic structure of G/hBN with twist angle 0.5032$%
^{\circ }$. (a), (b) and (c) are calculated with relaxation while (d), (e),
and (f) are without lattice relaxation. (a) and (d) display the band
structures; (b) and (e) are the density of states; (c) and (f) are the
conductance without magnetic field. In the band structure and DOS, the red
dashed lines denote the graphene monolayer results. In the conductance
(c) and (f), the red solid, blue dashed, and magenta dot lines are calculated
for device sizes 200nm$\times $200nm, 300nm$\times $300nm, and 400nm$\times $%
400nm, respectively.}
\label{Figs2}
\end{figure}

\begin{figure}[H]
\centering
\subfigure[]{%
\includegraphics[scale=0.7]{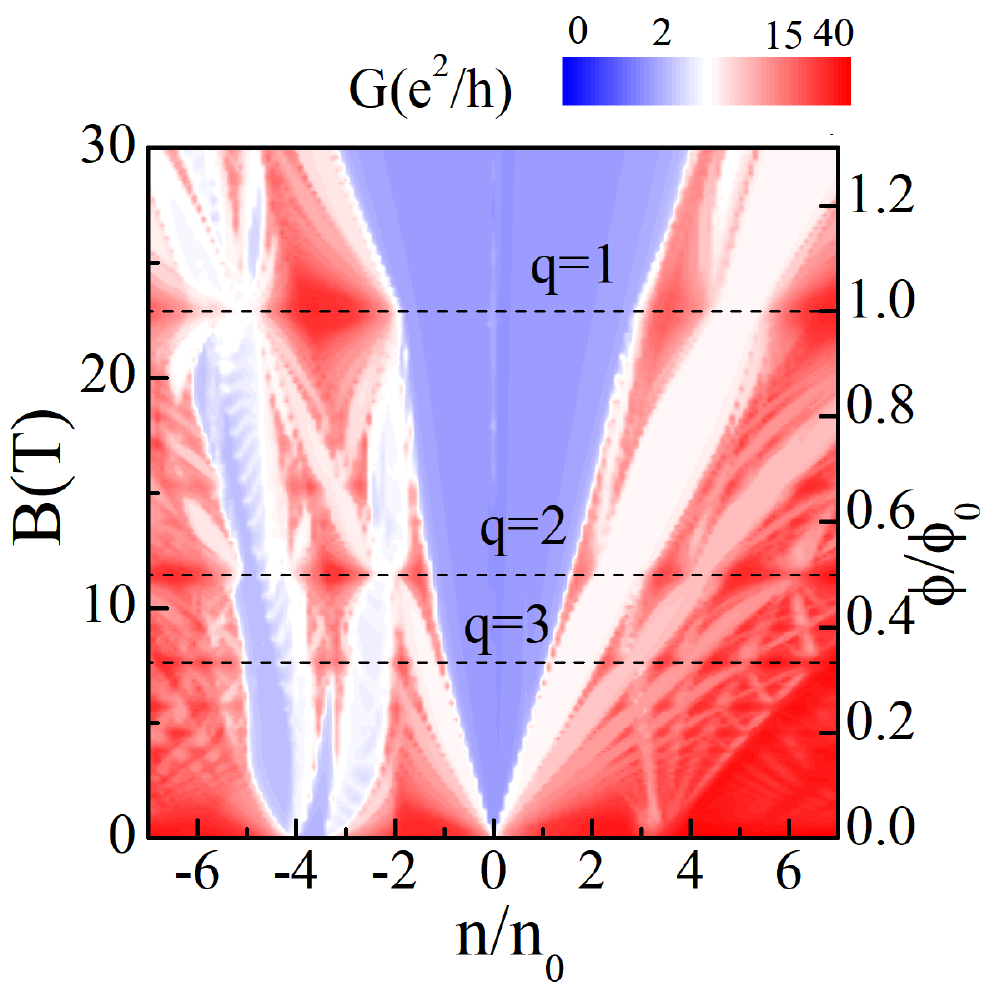}}
\hspace{6mm} \subfigure[]{%
\includegraphics[scale=0.7]{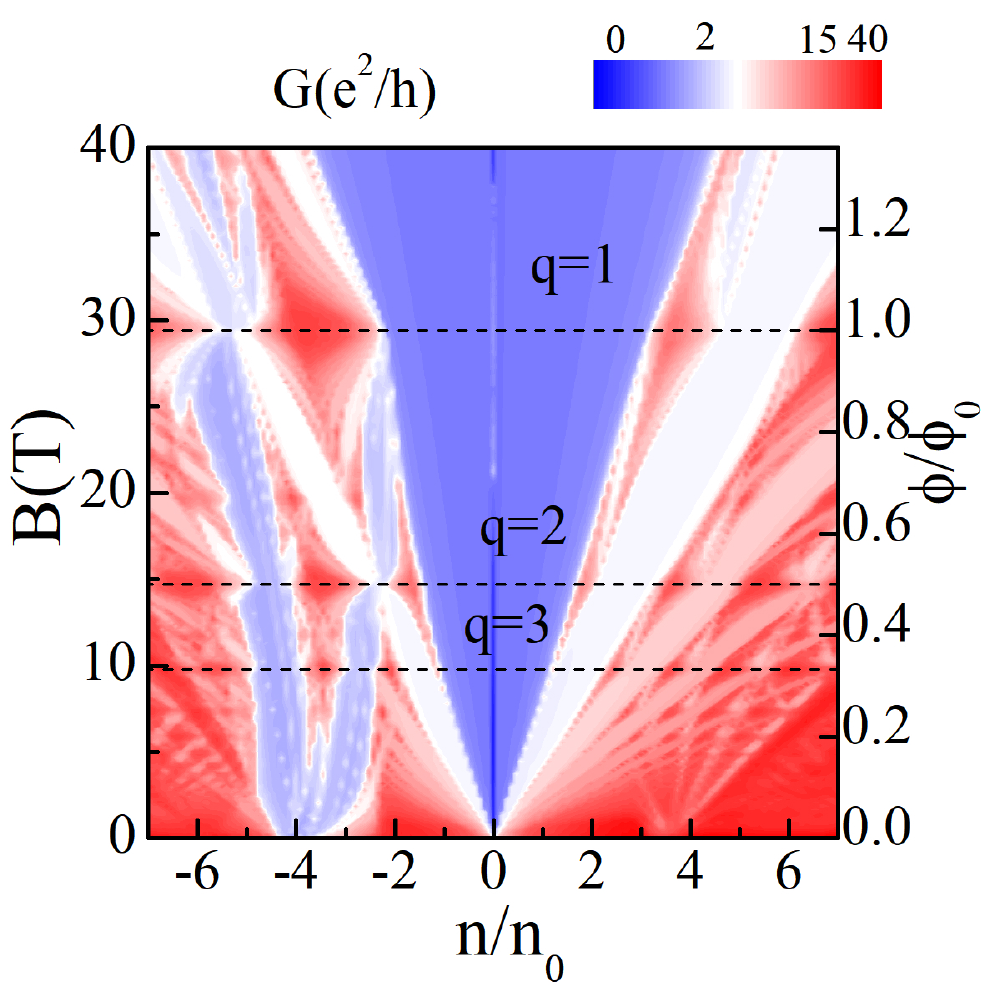}} %
\center\caption{Landau fans for the longitudinal conductance for twist angles (a) 0.0$^{\circ }$ and (b) 0.5032$^{\circ }$.}
\label{BS-141}
\end{figure}

\renewcommand\thetable{C\arabic{table}}
\renewcommand\thefigure{C\arabic{figure}}
\section{Appendix C: Edge dependence}
\setcounter{table}{0}
\setcounter{figure}{0}
\begin{figure}[H]
\center\includegraphics[scale=1]{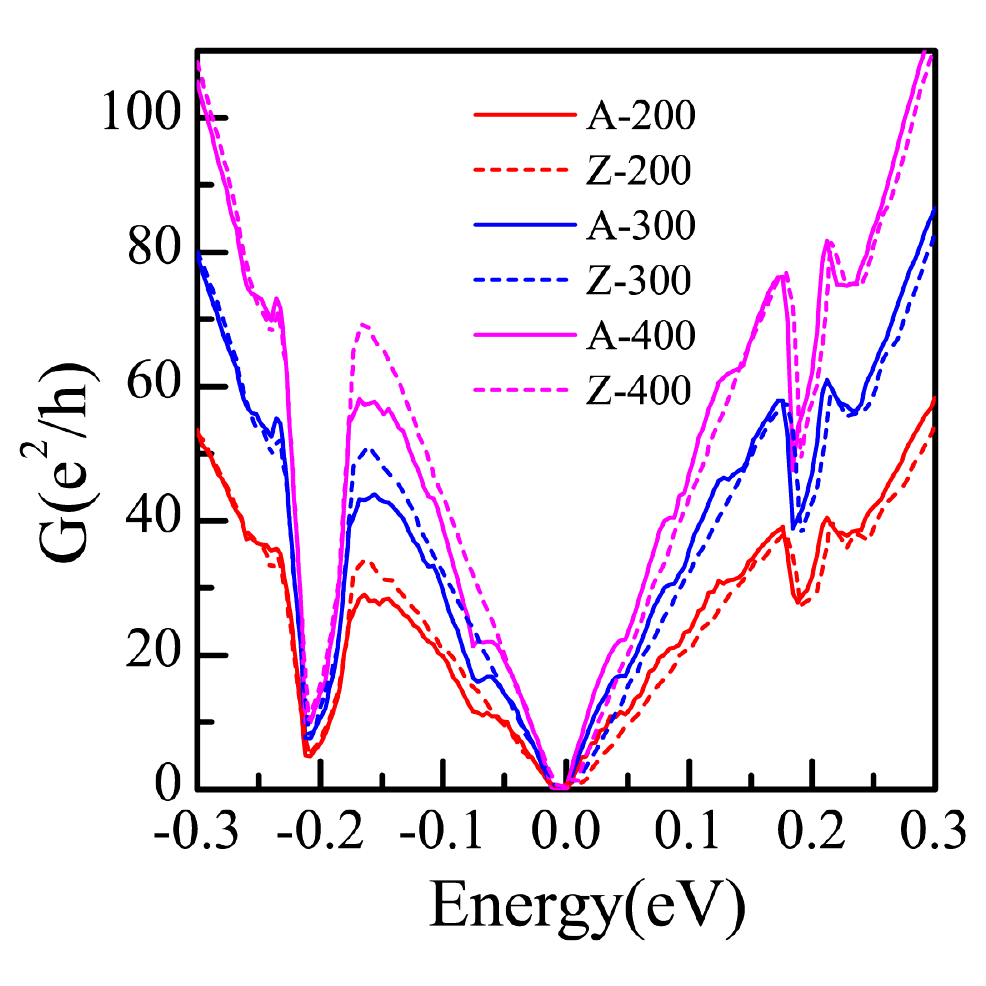}
\center\caption{The conductance of G/h-BN for devices with zigzag edge (Z) and armchair
edge (A), with twist angle 1.0047$^{\circ} $. Solid lines denote zigzag edges with  while dashed lines are for armchair edges. The device sizes are 200 nm $\times$ 200 nm, 300 nm$ \times $300 nm, and 400nm$\times $400nm, respectively.}
\label{Figs3}
\end{figure}

\renewcommand\thetable{D\arabic{table}}
\renewcommand\thefigure{D\arabic{figure}}
\section{Appendix D: Details of magnetoconductivity}
\setcounter{table}{0}
\setcounter{figure}{0}
\begin{figure}[H]
\centering
\center\includegraphics[scale=0.8]{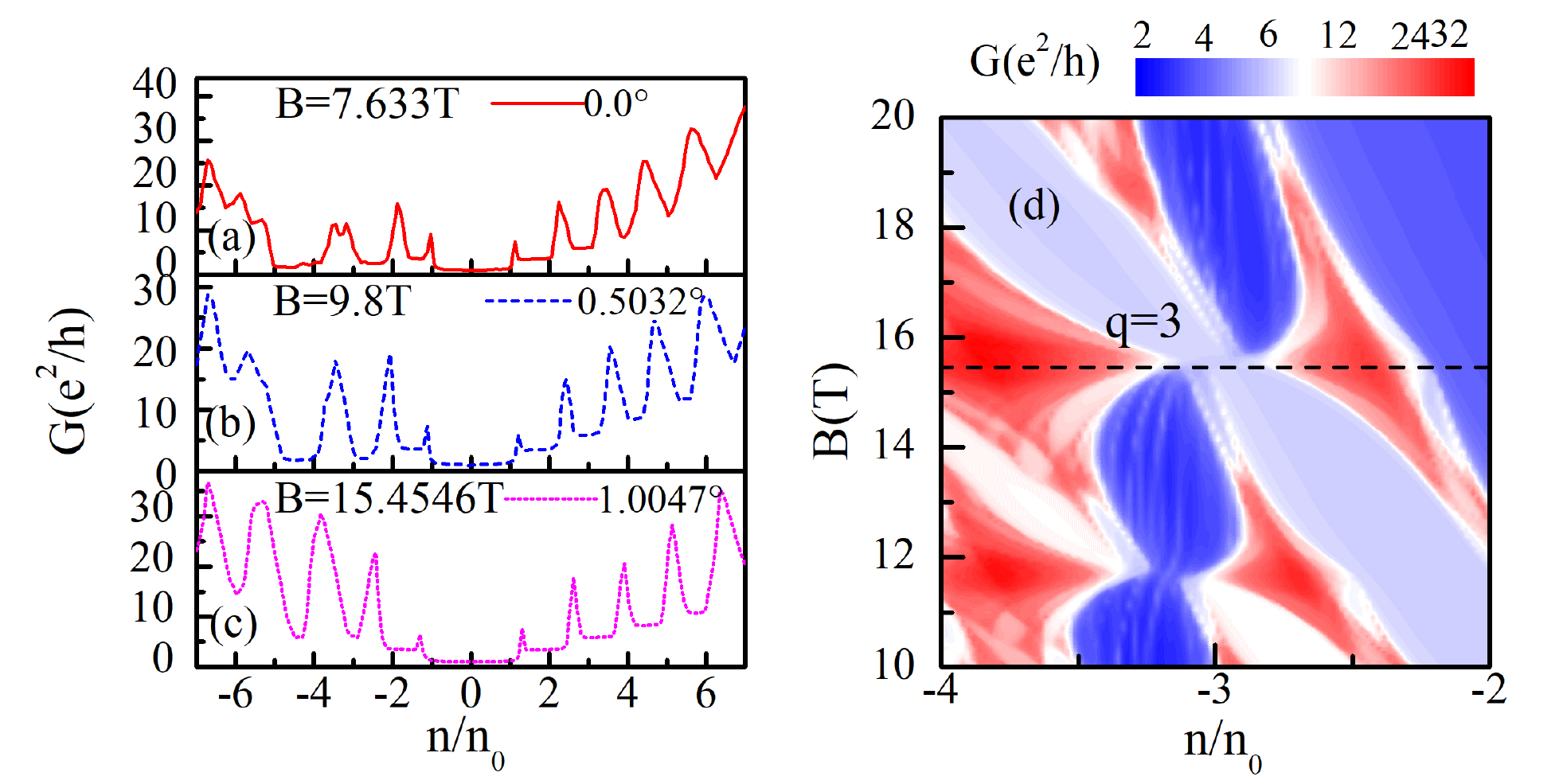}
\center\caption{The conductances along the $q=3$ line in Fig. \ref{fig2}, Fig. \ref{BS-141}(a) and Fig. \ref{BS-141}(b) for
magnetic fields B=7.633T, 9.8T, 15.4546T and the twist
angles = 0.0$^{\circ }$, 0.5032$^{\circ }$ and 1.0047$^{\circ }$.
(d) Enlarged view of magnetic Dirac point closing in Fig. 2 around the $q=3$ line.}
\label{Figs6}
\end{figure}

\renewcommand\thetable{E\arabic{table}}
\renewcommand\thefigure{E\arabic{figure}}
\section{Appendix E: Disorder effect}
\setcounter{table}{0}
\setcounter{figure}{0}
\begin{figure}[H]
\centering
\center\includegraphics[scale=0.8]{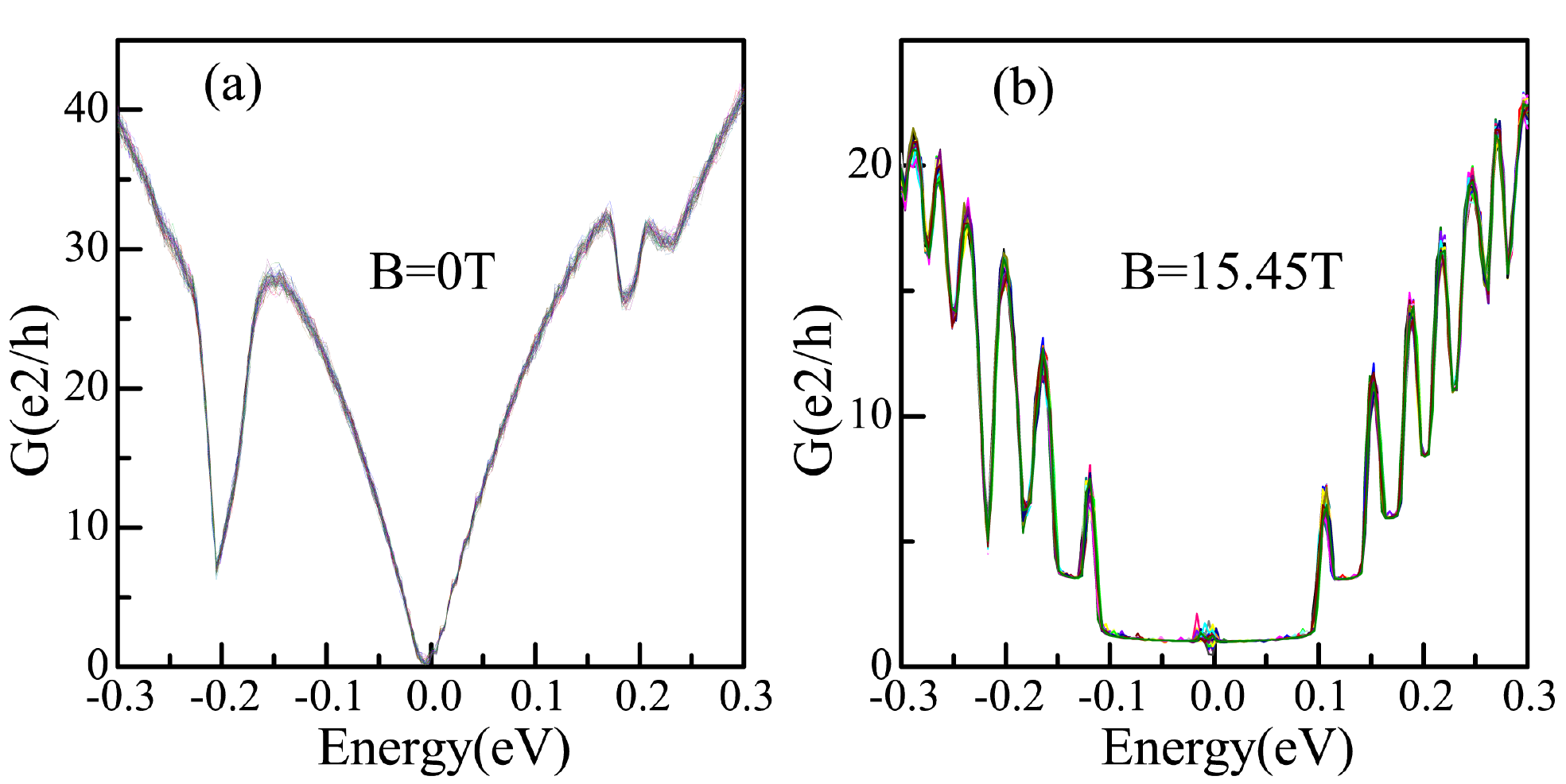}
\center\caption{Conductance for 60 different realizations of the disorder for a twist
angle = 1.0047$^{\circ}$ with and without magnetic field. The site-diagonal random potential $v_{i}$ varies within the range [-$V_{0}$, $V_{0}$], where $V_{0}$=0.5eV. The main features (e.g., the secondary Dirac points) stay unchanged, though there are small fluctuations in the magnitude of the conductance.}
\label{Figsdisorder}
\end{figure}

\renewcommand\thetable{F\arabic{table}}
\renewcommand\thefigure{F\arabic{figure}}
\section{Appendix F: Antidot lattice}
\setcounter{table}{0}
\setcounter{figure}{0}
\begin{figure}[H]
\centering
\center\includegraphics[scale=0.7]{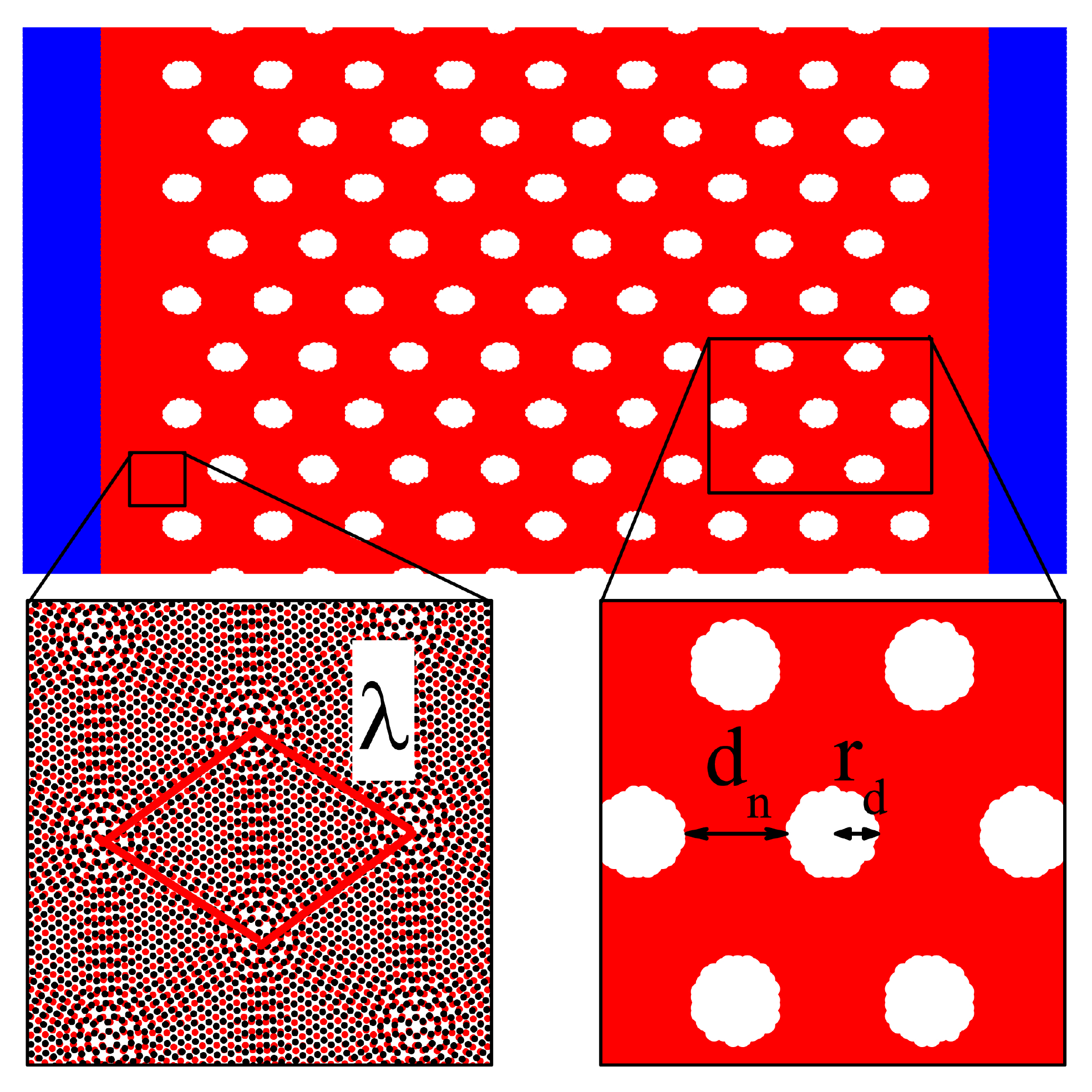}
\center\caption{Schematic of the device. Blue areas are the left and the right
lead, and red area is the scattering region. The enlarged figures (bottom
panels) show the moir{\'e} wave length $\protect\lambda $, the neck width $%
d_{n}$, and the radius $r_{d}$ of the antidot. The lattice constant of the
antidot lattice is $a_{antidot}=d_{n}+2r_{d} $. In our calculations we set $%
a_{antidot}=35$ nm, and vary the antidot's radius to generate different neck
widths.}
\label{Figs7}
\end{figure}
\end{widetext}

\end{document}